\documentstyle[prl,aps,12pt]{revtex}

\title{Protein folding, anisotropic collapse and blue phases}
\author{E. Pitard, T. Garel and H. Orland \\
Service de Physique Th\'eorique\\
CE-Saclay, 91191 Gif-sur-Yvette Cedex\\
France}
\date{\today}

\def\be{\begin{equation}}
\def\th0{\theta_0}

\def\bea{\begin{eqnarray}}
\def\ee{\end{equation}} 
\def\eea{\end{eqnarray}} 

\begin{document}
\maketitle
\vskip 1cm
\begin{abstract}
We study a homopolymer model of a protein chain, where each monomer
 carries a dipole moment. To mimic the geometry of the peptidic bond, 
 these dipoles are constrained to be locally perpendicular to the
 chain. The tensorial character of the dipolar interaction leads
 naturally to a (tensorial) liquid crystal-like order parameter. For
 non chiral chains, a mean field study of this model shows that a
 classical $\theta$ collapse transition occurs first;  at lower temperature,
 nematic order sets in. For chiral chains, an anisotropic (tensorial)
 collapse transition  may occur before the $\theta$ temperature is
 reached: the ordered  phase can be described as a ``compact phase of
 secondary structures'',  and possesses great similarities with the
 liquid crystal blue phases. 


\end{abstract}
\vskip 1cm
\noindent\mbox{Accepted for publication in:} \hfill 
\mbox{Saclay, SPhT/97-025}\\ \noindent \mbox{``J. de Physique I''}\\ 
\vskip 1cm\noindent \mbox{PACS: 61.41; 87.15.Da; 64.70.Md} \newpage

\newpage

\section{Introduction}
\label{sec: intro}
Proteins are chiral heteropolymers, made out of twenty species of
monomers (aminoacids). Natural proteins have the property of folding
into an (almost) unique compact native structure, which is of
biological interest \cite{Creighton84,Creighton92}. The compactness of
this unique native state is largely due to the existence of an optimal
amount of hydrophobic aminoacid residues \cite {Dill95}, since these
biological objects are usually designed to work in water. The
existence of secondary structures \cite{Pauling_Corey} plays a
predominant role in the existence of the unique native state. From a
theoretical point of view, one may either emphasize the heterogeneous
character of these biopolymers \cite{Garel}, or choose to focus on the
more crystalline-like secondary structures ($\alpha$-helices and
$\beta$-sheets). In the first approach, folding is identified to a
freezing transition; in the second approach, that we follow in this
paper, the folding transition is more like a gas-solid
transition. Previous studies along the latter path made use of
simplified lattice models \cite{Doniach}. We consider here a
microscopic continuous model of a homopolymer chain, where each
monomer carries a dipole moment. The rationale for this choice is
clear if one interprets the chemist's hydrogen bonds as the
interaction between
electrostatic dipoles (the secondary structures approximately
corresponding to the classic dipole equilibrium positions). The
dipolar moment of a $CONH$ is $p_0 \simeq 3.6$D, yielding a typical
dipolar temperature (temperature below which dipolar effects become
relevant) of $T_D ={p_0^2 \over 4 \pi \epsilon_0 \epsilon_r d^3}$
where $\epsilon_r$ is the relative dielectric constant of the medium,
and $d$ is a typical distance between dipoles. The electrostatics of
proteins in water is a somewhat controversial issue \cite{Honig}, so
we will content ourselves with an estimate of $T_D$ in the native
state. It is then appropriate to take $\epsilon_r \sim 5$
\cite{Honig}. With $d \sim 3.8$ \AA \cite{Creighton84}, we 
find $T_D \simeq {2000\over \epsilon_r} \simeq 400 K$ ,  so that
dipolar effects may play a crucial role in the energetics of the
folding (or unfolding) transition. The spatial range of the
dipole-dipole interaction can be finite or infinite: its  
angular dependence (or tensorial character) is the crucial issue of
this paper.  To make a closer connexion with proteins, we assume that
the dipole moments are constrained to be perpendicular to the
chain. In this way, a monomer can be thought of as linking two
successive $C_{\alpha}$ carbon atoms, and its dipole moment
represents the peptid $CONH$ dipole. A finite range of the dipolar
interaction may further describe the finite range character of the
hydrogen bonding 
interactions. The Hamiltonian of the model then reads  
\begin{equation}
\label{ham}
\beta{\cal H} = {v_0 \over 2}  \sum_{i \ne j} \delta(\vec r_{i}-\vec
r_{j}) + {\beta \over 2} \sum_{i \ne j} \sum_{\alpha, \gamma}
p_{i}^{\alpha} G_{\alpha \gamma}(\vec r_{i},\vec r_{j}) p_j^{\gamma}
\end{equation}
In equation (\ref{ham}), $\beta={1 \over T}$ is the inverse
temperature, $v_0$ is the excluded volume, $\vec r_{i}$ denotes the
spatial position of monomer $i$ ($i=1,2,...N$), and $\vec p_{i}$ its
dipole moment. If necessary, three body repulsive interactions may be
introduced, to avoid collapse at infinite density. The (infinite
range) dipolar tensor reads 
\begin{equation}
\label{dipo}
 G_{\alpha \gamma}(\vec r,\vec r')= A {1 \over \vert \vec r -\vec
 r'\vert^{3}} (\delta_{\alpha \gamma}-3 v_{\alpha}v_{\gamma})
\end{equation}
with $v_{\alpha}={(\vec r-\vec r')_{\alpha} \over \vert \vec r -\vec
 r'\vert}$ and $A$ is a prefactor containing the dielectric constant
 of the medium.
The dipolar interaction (\ref{dipo}) is cut-off at small distances
 ($|\vec r-\vec r'|< a$) and may also be cut-off at large distances
 by an exponential prefactor. The partition
 function of the model (\ref{ham}) is given by:
\begin{equation}
\label{partit}
{\cal Z} =\int \prod_{i} d {\vec r_i} d {\vec p_i} \
{\delta(\vert{\vec r_{i+1}-{\vec r_{i}}}\vert-a)} \ \delta (\vert{\vec
p_{i}}\vert -p_0) \  \delta( \vec p_i \cdot (\vec r_{i+1}-\vec r_{i}))
 \ \exp \left( - \beta  
{\cal H} \right)
\end{equation}
In equation (\ref{partit}), $a$ denotes the Kuhn length of the
monomers, and $p_0$ is the magnitude of their dipole moment.  The
third $\delta$-function restricts the dipole moment 
to be perpendicular to the chain; this mimics the free rotation of the
$CONH$ dipole around the $C_{\alpha}-C_{\alpha}$ ``virtual bond''.
As it stands, this partition
function is not easy to evaluate. In particular the constraint of
fixed length dipoles is not easy to implement. We have therefore
studied the constrained dipolar chain in (i) in the framework of the
virial expansion, with fixed length dipoles (ii) in the framework of a
Landau theory, where we replace the fixed dipole length by a Gaussian
distributed length.
\section{The virial approach}
\label{sec: virial}
Since the virial extension is quite familiar in the field of polymeric
liquid crystals \cite{Vroege_Lekk}, we will be rather sketchy in our
presentation. Starting from equations (\ref{ham}-\ref{partit}), we
write
\newpage
\bea
\label{viri}  
\exp\left(-{\beta} \sum_{i <j} \sum_{\alpha, \gamma}
p_{i}^{\alpha} G_{\alpha \gamma}(\vec r_{i},\vec r_{j})
p_j^{\gamma}\right) &=& \prod_{i < j} (1+f_{ij}) \nonumber \\
 &\simeq & 1+\sum_{i<j}f_{ij}  + \dots \nonumber \\
&\simeq & e^{ \sum_{i<j}f_{ij}  + \dots}
\eea
where $ f_{ij} = f(\vec r_i - \vec r_j) = \exp\left({- \beta {\vec p_i}\cdot {\bf G}_{ij}\cdot {\vec
p_j}}\right) -1 $.
To lowest non trivial order in $f$, we get:
\bea
\label{viri2}
{\cal Z} &=&\int \prod_{i} d {\vec r_i} 
{\delta(\vert{\vec r_{i+1}-{\vec r_{i}}}\vert-a)}
 \ \exp \left( - \beta {\cal H}_{pol}(\{\vec r_i\})\right)\nonumber \\
&\times& \exp \left(\sum_{i<j} \int  d\vec p_i d\vec p_j  f_{ij}
\delta (\vert{\vec
p_{i}}\vert -p_0) \  \delta( \vec p_i \cdot (\vec r_{i+1}-\vec r_{i}))
\delta (\vert{\vec
p_{j}}\vert -p_0) \  \delta( \vec p_j \cdot (\vec r_{j+1}-\vec r_{j}))\right)
\eea
where $\beta {\cal H}_{pol} = {v_0 \over 2} \sum_{i \ne j}
\delta (\vec r_i - \vec r_j)$ .
Introducing the joint distribution function:
\be
\label{joint}
g(\vec r, \vec u) = \sum_{i} \delta(\vec r - \vec r_i)\ 
\delta(\vec u - \vec u_i)
\ee
where $\vec u_i = (\vec r_{i+1} - \vec r_i)/a $, the partition function
(\ref{viri2}) reads:
\bea
\label{viri3}
{\cal Z} &=& \int {\cal D}g {\cal D} \hat g \exp \left( i \int d\vec r
d\vec u \ {\hat g }g - {v_0 \over 2}  \int d\vec r (\int d\vec u \ g)^2 \right)
\nonumber \\
&\times& \exp\left({1\over2} \int d\vec r d\vec r' d\vec u d\vec u' g (\vec
r, \vec u) V(\vec r, \vec u |\vec r', \vec u' ) g (\vec
r', \vec u')
+ \log \zeta ( \hat g)
\right)
\eea
where 
\be
\label{V}
V(\vec r, \vec u |\vec r', \vec u' )= \int d\vec p d\vec p' 
\delta (|{\vec p}| -p_0) \  \delta( \vec p \cdot \vec u)
\delta (|{\vec p'}| -p_0) \  \delta( \vec p' \cdot \vec u')
f (\vec r - \vec r')
\ee
and $\zeta$ is the polymeric partition function given by:
\be
\label{zeta}
\zeta(\hat g) = \int \prod_{i} d {\vec r_i} \ 
{\delta(\vert{\vec r_{i+1}-{\vec r_{i}}}\vert-a)} 
e^{- i \sum_i \hat g (\vec r_i, \vec u_i)}
\ee
In the spirit of mean field or variational approaches
\cite{Vroege_Lekk}, we look for a homogeneous (space independent)
distribution function: $g(\vec r, \vec u) = \rho \phi (\vec u)$, where
$\rho = {N \over \Omega}$ is the monomer concentration and the angular
distribution function $\phi$ is normalized to $1$. The free energy per
monomer reads:
\be
\label{free1}
\beta {\cal F} = \int d\vec u \phi(\vec u) \log (\phi(\vec u)) + {v_0 \over
2} \rho - {1 \over 2} \ \rho \int d\vec u d\vec u' \phi(\vec u) {\cal V} (\vec u \cdot
\vec u') \phi(\vec u') + O({\rho}^2)
\ee
where 
\be
{\cal V}(\vec u \cdot \vec u') = 
\int d \vec r d\vec p d\vec p' 
\delta (|{\vec p}| -p_0) \  \delta( \vec p \cdot \vec u)
\delta (|{\vec p'}| -p_0) \  \delta( \vec p' \cdot \vec u')
f (\vec r)
\ee
and the $O({\rho}^2)$ term in equation (\ref{free1}) represents the 
three body repulsive interactions.
Using the Onsager variational form \cite{Onsager} 
\be
\phi(\theta) =
{\alpha  \over 4 \pi \sinh \alpha} \cosh (\alpha \cos \theta) 
\ee
where $\theta$ is the polar angle of the unit vector $\vec u$, we obtain
\bea
\label{free2}
\beta {\cal F} =&&  \log ({\alpha \over {4\pi e \tanh {\alpha}}}) + {2
\over \sinh \alpha} (\arctan(e^{\alpha}) - {\pi \over 4})
\nonumber\\
&& + {v_0 \over
2} \rho \left(1-{1 \over t^2}-{3 \over 80 t^2} ((1-2w(\alpha))^2+2w^{2}(\alpha)-{1
\over 3})\right) + O({\rho}^2)
\eea
where $t= {T \over T_{\theta}}$, with $T_{\theta} = {2 \over 3}
\sqrt{a^3 \over v_0} T_D$, and $w(\alpha)=({\coth \alpha \over \alpha}-{1
\over {\alpha}^2})$. We have also assumed that $d \sim a$, i.e. that an
estimate of the distance between dipoles is given by the monomer
length. Minimizing equation (\ref{free2}) with respect to $\alpha$ and
$\rho$ yields the following results: 
\begin{itemize}
\item{i)} A usual second order $\theta$-like collapse transition to a finite
density phase occurs at temperature $t=1$. Since $v_0 \sim a^3$, we
have $T_{\theta} \sim T_D$.
\item{ii)} In the dense phase, a discontinuous nematic-like transition
occurs at temperature $t_N = {T_N \over T_{\theta}} \simeq  {1 \over 3}$,  with a small density
discontinuity. Since $T_N$ is a variational estimate, the true
transition temperature is necessarily higher.
\end{itemize}
These results show that dipolar interactions are responsible for both
transitions. We remind the reader 
that they are expected to be correct at small monomer concentration,
since they rely on the virial expansion.
At higher monomer concentration, one may find other
phases (e.g. crystalline phases \cite{Groh_Die}). 
An extension of this approach to chiral chains is possible
\cite{Straley}. In this case, one expects to encounter a phase
transition towards an inhomogeneous phase (e.g. cholesteric) at a
temperature $T_H$  higher than $T_N$. The possibility of a crossing
between $T_{\theta}$ and $T_H$  must therefore be considered. As shown
in the next section, the Landau approach to the constrained dipolar
chain provides a more convenient framework to study this possibility. 

\section{Landau theory}
\label{sec: Landau}
\subsection{Formal developments}
\label{sec:formal} 

At a qualitative level, experimental phase diagrams are usually in
agreement with the Landau theory of phase transitions. Roughly
speaking, this theory, which rests over the use of symmetry arguments,
may be summarized as follows 
\begin{itemize}
\item(i)  one looks for an order parameter $Q(\vec r)$ that describes
the ordered phase(s) at point $\vec r$. 
\item(ii) one builds a free energy density as an expansion in $Q(\vec
r)$, and in the first spatial derivatives of $Q(\vec r)$.   
\item(iii) one minimizes the total free energy with respect to $Q(\vec
r)$.
\end{itemize}
This theory clearly assumes that the order parameter $Q(\vec r)$
varies on distances much larger than the microscopic length
scales. Following this line, we first soften the constraint of fixed
length dipoles in equation (\ref{partit}) and replace it by a Gaussian
constraint (the replacement  of fixed length by Gaussian spins is
familiar in the theory of phase transitions). We therefore have
\begin{equation}
\label{partit2}
{\cal Z}_{L} =\int \prod_{i} d {\vec r_i} d {\vec p_i} \
{\delta(\vert{\vec r_{i+1}-{\vec r_{i}}}\vert-a)} \ {1 \over ({2
\pi p_0^2})^{3/2}} \ e^{-{{\vec p_i}^2 \over {2p_0^2}}} \  \delta( \vec p_i \cdot (\vec r_{i+1}-\vec r_{i}))
 \ \exp \left( - \beta  
{\cal H} \right)
\end{equation}
where the subscript {L} on the partition function stands for
Landau and the Hamiltonian ${\cal H}$ is given by equation (\ref{ham}). 
Using the identity
\be
\label{identi}
\delta(\vec y)=\lim_{\lambda\to \infty}\left(\lambda \over {2\pi}\right)^{3/2}
e^{-{\lambda{\vec y}^2 \over 2}}
\ee
we may now perform the (Gaussian) integrals over the dipole moments
$\vec p_i$ in equation (\ref{partit2}). As a result, the problem now depends
only on the polymeric degrees of freedom. Introducing
the tensorial (order) parameter $Q_{\alpha\mu}(\vec r)$ by  
\be
\label{paramor}
Q_{\alpha\mu}(\vec r)= \sum_{i}\left({(\vec u_i)_{\alpha}(\vec
u_i)_{\mu}}-\delta_{\alpha \mu}\right) \ \delta(\vec r-\vec r_{i})
\ee
where the notation $\vec u_i = (\vec r_{i+1} - \vec r_i)/a$ was again
used, we get
\bea
\label{partit3}
{\cal Z}_{L} =&& \int \prod_{(\alpha,\mu),\vec r} {\cal D}Q_{\alpha
\mu}(\vec r) \ \delta\left(Q_{\alpha\mu}(\vec r)- \sum_{i}({(\vec
u_i)_{\alpha}(\vec u_i)_{\mu} }-\delta_{\alpha \mu}) \
\delta(\vec r-\vec r_{i})\right)\nonumber\\
&&
\int \prod_{i} d {\vec r_i}  \ {\delta(|{\vec
r_{i+1}-{\vec r_{i}}}|-a)} \ \exp \left( - \beta  
{\cal H}_{pol}\right) \ \exp \left(- {1 \over 2} {\rm Tr} \log {\bf B}\right)
\eea
where the matrix ${\bf B}$ is defined by its matrix elements
\be
\label{partit4}
B_{\alpha \gamma}(\vec r,\vec r')=
\delta_{\alpha\gamma}\delta(\vec r-\vec r')+\beta p_0^2
\sum_{\mu}Q_{\alpha\mu}(\vec r) G_{\mu \gamma }(\vec r,\vec r')
\ee
Using the identity 
\be
\label{delta1}
\delta\left(Q_{\alpha\mu} -d_{\alpha \mu}\right)=\int {{\cal D} \hat
Q_{\alpha\mu} \over 2\pi} \exp {i\hat
Q_{\alpha\mu}\left(Q_{\alpha\mu}-d_{\alpha \mu}\right)}
\ee
one may write
\bea
\label{partit5}
{\cal Z}_{L} =&& \int \prod_{(\alpha,\mu),\vec r} 
{\cal D}Q_{\alpha
\mu}(\vec r){\cal D}\hat Q_{\alpha\mu}(\vec r) 
\exp(i \int d \vec r \sum_{\alpha \le \mu} Q_{\alpha \mu}(\vec r ) Q_{\alpha \mu}(\vec r )) 
\nonumber\\
&& \exp \left( - \beta {\cal H}_{pol}({\bf Q})\right) \ \exp \left(- {1
\over 2} {\rm Tr} \log {\bf B}\right) \exp(\log\zeta({\bf \hat Q}))
\eea
with
\be
\label{chaine1}
\zeta({\bf \hat Q})=\int \prod_{i} d {\vec r_i}  \ {\delta(|{\vec
r_{i+1}-{\vec r_{i}}}|-a)}\exp\left(-i\sum_{\alpha\le\mu}\sum_{i}\hat
Q_{\alpha\mu}(\vec r_{i})({(\vec u_i)_{\alpha}(\vec u_i)_{\mu}
}-\delta_{\alpha \mu})\right)
\ee 
and where we have expressed ${\cal H}_{pol}$ as a function of ${\bf
Q}$ (see equation (\ref{rho}) below). So far, no approximations beside the assumption of a Gaussian distributed 
length of the dipoles were made. We stress the fact that the number of
independent variables $\hat Q_{\alpha \mu}(\vec r)$ is equal to six,
since ${\bf Q}$ is a symmetric rank 2 tensor. This is why the sum over
the components $(\alpha,\mu)$ in equations (\ref{partit5}, \ref{chaine1}) is
restricted to $\alpha\le\mu$. 

\subsection{The physical order parameters} 
\label{sec: orderpa}
At this stage, it may be useful to establish the relation of the
previous section with the virial approach. It is clear that the tensor
(or matrix) ${\bf Q}(\vec r)$ can also be expressed through the joint
distribution function of equation (\ref{joint}) as 
\be
\label{paramor2}
Q_{\alpha\mu}(\vec r)=\int d\vec u \ g(\vec r, \vec u) \left({(\vec
u)_{\alpha}(\vec u)_{\mu} }-\delta_{\alpha \mu}\right) 
\ee
Furthermore, one may build from ${\bf Q}(\vec r)$ a (scalar) density
\be
\label{rho}
\rho(\vec r)= -{1 \over 2} \ {\rm Tr} \ {\bf Q}(\vec r)
\ee
 and a
traceless tensor 
\be
\label{q}
{\bf q}(\vec r)={\bf Q}(\vec r)-{{\bf I} \over 3}{\rm Tr} \
{\bf Q}(\vec r)
\ee
where ${\bf I}$ is the unit tensor.
Assuming that both quantities are space
independent, we easily recover the results of section \ref{sec: virial}.
For space dependent order parameters, the Landau-Ginzburg free energy
density ${\cal F}_L (\vec r)$ is defined by:
\be
\label{partit6}
{\cal Z}_L= \int {\cal D} \rho(\vec r)
{\cal D} q_{\alpha \mu}(\vec r) e^{- \beta \int d\vec
r {\cal F}_L(\vec r)}
\ee
It can be evaluated 
by 
\begin{itemize}
\item{(i)} performing a saddle point method on ${\bf \hat Q}$ in
equation (\ref{partit5}) and replacing ${\bf \hat Q}$ as a function of
$\bf Q$.
\item{(ii)} expanding the resulting free energy density in the
exponent of (\ref{partit5})
in powers of
$\bf Q$ and performing a local gradient expansion. 
Note that all terms of the expansion could have been obtained
from symmetry considerations.
\item{(iii)} expressing the free energy density in terms of $\rho(\vec
r)$ and ${\bf q}(\vec r)$.
\end{itemize}    

The resulting Landau free energy density reads:
\bea
\label{landau}
{\cal F}_L(\vec r) &=& {a_0^2 \over 2} (\nabla \rho (\vec r))^2+
{a_2 \over 2}(T-T_{\theta}) \rho^2 (\vec r) +
{a_3 \over 3} \rho^3 (\vec r) + \cdots \nonumber \\ 
&+& {b_0 \over 2} \partial_{\alpha} q_{\alpha \mu} (\vec r)
\partial_{\alpha'} q_{\alpha' \mu} (\vec r) +{b_0' \over 2}
\partial_{\gamma} q_{\alpha \mu} \partial_{\gamma} q_{\alpha \mu} +
{b_2 \over 2
}(T-T_0) {\rm Tr} \ {\bf q}^2 (\vec r) \nonumber \\
&-& {b_3 \over 3!} {\rm
Tr}\ {\bf q}^3 (\vec r) + {b_4 \over 4!}\left({\rm Tr}\ {\bf q}^2(\vec
r)\right)^2 + \cdots \nonumber \\
&+& {c_0 \over 2} \rho(\vec r) \partial^2_{\alpha \mu} q_{\alpha
\mu}(\vec r) - {c_1 \over 2} \rho(\vec r)  {\rm Tr} \ {\bf q}^2 (\vec
r)
+ \cdots
\eea
where we use the standard summation over repeated indices. The various
coefficients $a, b, c$ in (\ref{landau}) can in principle be
calculated using the method described above. In particular, the
temperature $T_{\theta}$ is the same as that defined in section
\ref{sec: virial} and $T_0$ would be the nematic transition
temperature if there was no cubic term in ${\bf q}$ in equation (\ref{landau}).
\subsection{Including chirality}
\label{sec: chiral}
The observation that the backbone of proteins has chirality can be
appreciated through the Ramachandran plot \cite{Creighton84}. The
chirality of the $C_{\alpha}$ ``virtual chain'' is 
predominantly associated with the $\phi$ dihedral angles. It is not easy to
take chirality into account at a microscopic level \cite{Harris},
while it is so at a Landau-like level \cite{PGG}, since then, only
very general symmetry considerations matter. In particular, if the
system is described by a tensor ${\bf q}(\vec r)$, the absence of
inversion symmetry ensures that a new (scalar) invariant shows up in
equation (\ref{landau}), namely 
\be
\label{chiral}
{\cal F}_{chiral}(\vec r)= D \
\epsilon_{\alpha\mu\nu}q_{\alpha\delta}\partial_{\mu}q_{\nu\delta} 
\ee
where $D$ is a measure of the strength of the chirality and
$\epsilon_{\alpha\mu\nu}$ the antisymmetric third rank tensor. The
full Landau free energy density ${\cal F}_L^{tot}(\vec r)$ of the
constrained dipolar chiral chain then reads
\be
\label{full}
{\cal F}_L^{tot}(\vec r)={\cal F}_L(\vec r)+{\cal F}_{chiral}(\vec r)
\ee
where ${\cal F}_L(\vec r)$ and
${\cal F}_{chiral}(\vec r)$ are given respectively in equations (\ref{landau})
and (\ref{chiral}). It is clear that equation (\ref{full}) describes
the coupling between a usual $\theta$ collapse described by $\rho(\vec
r)$ and blue phase-like ordering \cite{Mermin} described by ${\bf
q}(\vec r)$. For our purposes, it is important to note that equation
(\ref{chiral}) implies that a modulated order in ${\bf
q}(\vec r)$ sets in at a temperature $T_{\chi}$ which is higher
\cite{Mermin,Grebel} than the
temperature $T_N$ of the uniform (nematic) order; on the other hand equation
(\ref{landau}) shows that, to lowest order, the $\theta$ collapse occurs at 
the same $D$-independent temperature $T_{\theta}$. Before exhibiting the
crossing of the two temperatures $T_{\chi}$ and $T_{\theta}$ in a
specific case, we briefly summarize the existing state of the art for
blue phases.   
\subsection{Summary of the blue phases folklore}
\label{sec: blueph}
In chiral systems, the blue phases (BP) may show up in a narrow temperature
range ($\le 1$K) between the simple
cholesteric (helical) liquid cristal phase and the isotropic
liquid. Their theory is quite complicated \cite{Mermin,Grebel}; their
ordering is described by a symmetric rank 2 tensor which can, in some
appropriate limits, be linked to the molecular orientations.  So far
two cristalline phases (BP I, BP II) and one non-cristalline phase
(BP III) have been found. More complicated phases may exist
when an electric field is present. 

The cristalline phases are
characterized by a body centered (BP I) or simple cubic (BP II)
lattice. The unit cell contains of the order of $10^7$ molecules, so
that the Landau expansion may be a priori trustworthy. Note that these
cristalline structures are threaded by disclinations (line defects of
the molecular orientations). 

The non cristalline phase (BP III) is of
particular interest \cite{Lubensky}. It has the same macroscopic
symmetry as the isotropic liquid, and the (BP III)-(isotropic liquid)
phase transition has a critical end point, much like the liquid-gas
transition. Furthermore, attempts have been made to describe this
phase as resulting from an unstable localized mode (see the references
in \cite{Mermin}), as opposed to an extended mode for the cristalline
BP's.   
\section{A simple example of anisotropic collapse}
\label{sec: simple}
We will exhibit, in a simple way, the crossing between the $\theta$
(uniform) collapse temperature $T_{\theta}$ and the temperature
$T_{\chi}$ below  which a modulated order sets in.
Following references \cite{Grebel,Brazo}, we consider in equation
(\ref{full}) the appearance of helicoidal order in the tensor ${\bf
q}(\vec r)$. At this point, it is perhaps useful to recall the reader
that the various ``helices'' of this section are not the
$\alpha$-helices present in proteins. Indeed, these ``helices'' are
linked to the order parameter of equation (\ref{q}), and, as in
all polymeric problems, it is difficult to deduce the chain
configuration(s) from the physical order parameter(s). The ${\bf
q}(\vec r)$ helicoidal order is described by: 
\be
\label{helixpara}
q_{\alpha\mu}(\vec r)=q_{\alpha\mu}(z)={1 \over \sqrt{6}} \ q\sin
{\phi}  \ M_{\alpha\mu}+{1 \over \sqrt2} \ q\cos{\phi} \ N_{\alpha\mu}(z)
\ee
where the matrices ${\bf M}$ and ${\bf N}(z)$ are given by
\be
\label{matr1}
{\bf M}=\pmatrix{{1}&{0}&{0}\cr
                {0}&{1}&{0}\cr
                {0}&{0}&{-2} \cr}
\ee
and
\be
\label{matr2}
{\bf N}=\pmatrix{{\cos k_0z}&{\sin k_0z}&{0}\cr
                {\sin k_0z}&{-\cos k_0z}&{0}\cr
                {0}&{0}&{0} \cr}
\ee
where $\vec k_0 ={D \over b_0'} \vec e_z$ is the wave vector of the modulation.
Usual cholesteric order (i.e. the ``uniaxial'' helix) corresponds to
$\phi={\pi \over 6}$, while the purely biaxial helix is described by
$\phi={0}$. We insert this trial order parameter
in equation (\ref{full}) and get:
\bea
\label{helic}
{\cal F}_L^{tot}(\vec r)&=& {a_0^2 \over 2} (\nabla \rho (\vec r))^2+
{a_2 \over 2}(T-T_{\theta})\  \rho^2 (\vec r) +
{a_3 \over 3} \rho^3 (\vec r) + \cdots \nonumber \\ 
&+& \left({b_2 } (T-T_0)- {c_1 }\ \rho(\vec r)\right)\ {q^2 \over 2}-{b_3
\over \sqrt 6 }\ \sin {3 \phi}\ {q^3 \over 6} + {b_4 }\ {q^4\over 24}
 + \cdots \nonumber \\
&-& {D^2\over 2 b_0'}\ q^2 \ \cos^2 \phi
\eea
This free energy density is to be minimized with
respect to  $q, \phi$ and $\rho(\vec r)$. 
Of particular importance are
the terms which couple the order parameters in
(\ref{full}).  With our choice of the order parameter
$q_{\alpha\mu}(z)$, the first non zero term
corresponds to the $c_1$ coefficient of equation
(\ref{landau}). Moreover, on physical grounds, this
coefficient must be positive, since an increase of the
density is expected to favor the appearance of a
liquid crystal type of ordering. Furthermore, the
order in $\rho(\vec r)$ can be shown to be uniform
($k=0$ ordering). 
Due to the large number of coefficients in the free energy density, it is
appropriate to work with dimensionless quantities, and we shall follow
 the notations of reference 
\cite{Grebel}.
The rescaled total free energy per unit volume $f$ reads:
\bea
\label{rescaled}
f &=& {1\over 4} (t - \alpha c) Q^2 -{1\over 4} {\kappa}^2 Q^2 \cos^2
\phi \nonumber \\
&-& Q^3 \sin 3 \phi + Q^4 + {1\over 2} t' c^2 + c^3
\eea
with
\bea
\label{def}
\hskip 2cm 
q &=& { 4\over \sqrt 6}\ {b_3 \over b_4}\ Q, \hskip 2cm
\rho = {2\over 3}\ {b_3 ^{4/3} \over b_4 a_3^{1/3}}\ c, \\
t &=&{1 \over \delta}(T-T_0),\hskip 2cm
t' = \gamma (T-T_{\theta}),\\
\gamma &=&{3\over 2}\ {a_2 b_4 \over a_3^{2/3} b_3^{4/3}}, \hskip 2cm
\delta = { b_3^2 \over 18\ b_2 b_4} \\
\kappa &=& 18\ k_0^2\ {b_4 b_0' \over b_3^2}, \hskip 2cm
c_1 = {b_3^{2/3}a_3^{1/3} \over 12} \ \alpha,
\eea
and
\be
{\cal F}_L^{tot} = \left({3 \over 2}\right)^3 {b_4^3 \over b_3^4} f,
\ee
We just quote here the conclusions of a study of equation (\ref{rescaled}).
Technical details will appear elsewhere \cite{Pitard}. 
The phase diagram can be studied in the (temperature, chirality) plane.
For high enough
values of the chirality parameter $\kappa$, 
\be
\kappa^2 \left( 1 - {\alpha^2 \over 32 \gamma (\delta \kappa^2 -
(T_{\theta} - T_0))} \right) > 9
\ee
we get a continuous
transition towards a biaxial helix at a reduced temperature
$t_{\chi}=\kappa^2$ corresponding to a real temperature
\be 
T_{\chi} = T_0 + \delta \kappa^2
\ee
Below the transition, the reduced density grows as $c
\simeq Q^2$.  The low temperature phase can be viewed as a compact
phase with helical order. From a protein-oriented point of view, this would
correspond to a direct transition from a coil state (denatured) to a
native state (i.e. compact with secondary structures).

At much lower chirality, when $T_{\chi} < T_{\theta}$, 
the scenario is more like
the one in section \ref{sec: virial}: there is first a $\theta$
collapse transition at $t_{\theta}'=0$, followed at a lower
temperature by a discontinuous transition towards a ``uniaxial''
helix. 

At intermediate chiralities, several phase diagrams are possible,
depending on the relative values of the parameters. In particular,
the location of the phase transition between the 
compact modulated phases (biaxial and
uniaxial) is very model dependent. 

In figure 1, we display a plausible phase diagram. 
\section{Conclusion}
\label{concl}
Using the fact that dipolar interactions imply the existence of a
tensorial order parameter (see equation (\ref{paramor})), we have
shown that chiral chains may have isotropic ($\theta$-like) or
anisotropic (blue phase -like) collapse transitions. This result was
found in a simple model with helicoidal order. The connexion with blue
phases is perhaps not very surprising since the molecules of some
of these compounds have dipoles perpendicular to the molecular axis. 
It is clear that the same questions that arise in the study of blue
phases should arise, 
mutatis mutandis, in the study of chiral dipolar chains:
\begin{itemize}
\item({i}) For very chiral systems, is the continuous transition
towards the compact biaxial phase preempted by a discontinuous transition
to a compact phase with a (hexagonal, body-centered
cubic,...) superposition  of biaxial helices?  
\item({ii}) Can one find a description of low chirality systems in
terms of double twist cylinders?
\item({iii}) Can one link the order in ${\bf q}(\vec r)$, the chain
configuration(s) and the dipole configuration(s)? Does a cristalline
order in ${\bf q}(\vec r)$ imply the existence of defects in the
orientation of the chain ? 

\end{itemize}

The application of the previous results to (bio)polymers can be
roughly summarized as follows. Since we are concerned with collapsed
phases with uniform density, we are typically in a melt situation. The
study of a single chain (such as a protein) is presently out of
reach. In both cases, a more detailed analysis
requires a better experimental knowledge of the various parameters
(chirality,...). Nevertheless, we think that our approach is
useful for the protein folding problem. In particular \cite{Pitard},
one may look for a double twist interpretation of
$\alpha$-helices and $\beta$-sheets, the possible existence of
localized instabilities (such as the ones in BP III \cite{kugler}) and
their implications for the folding process. Along these lines, it is
reassuring (although not completely unexpected) to note that the secondary
structures of real proteins are well characterized by a ${\bf q}(\vec
r)$ tensorial order parameter \cite{Delarue}.

We thank Marc Delarue, Jean-Renaud Garel, Roland Netz and Pawe\l \
Pieranski for pleasant and stimulating discussions.
\newpage
{\bf Figure caption}: Schematic phase diagram of the model in the
(temperature, chirality) plane. The phases are labelled by $C$ (coil),
$IG$ (isotropic globule), $AG_1$ and $AG_2$ (anisotropic globules). The
$C-IG$ transition, and the $C-AG_2$
transition (for large enough chirality), are continuous.

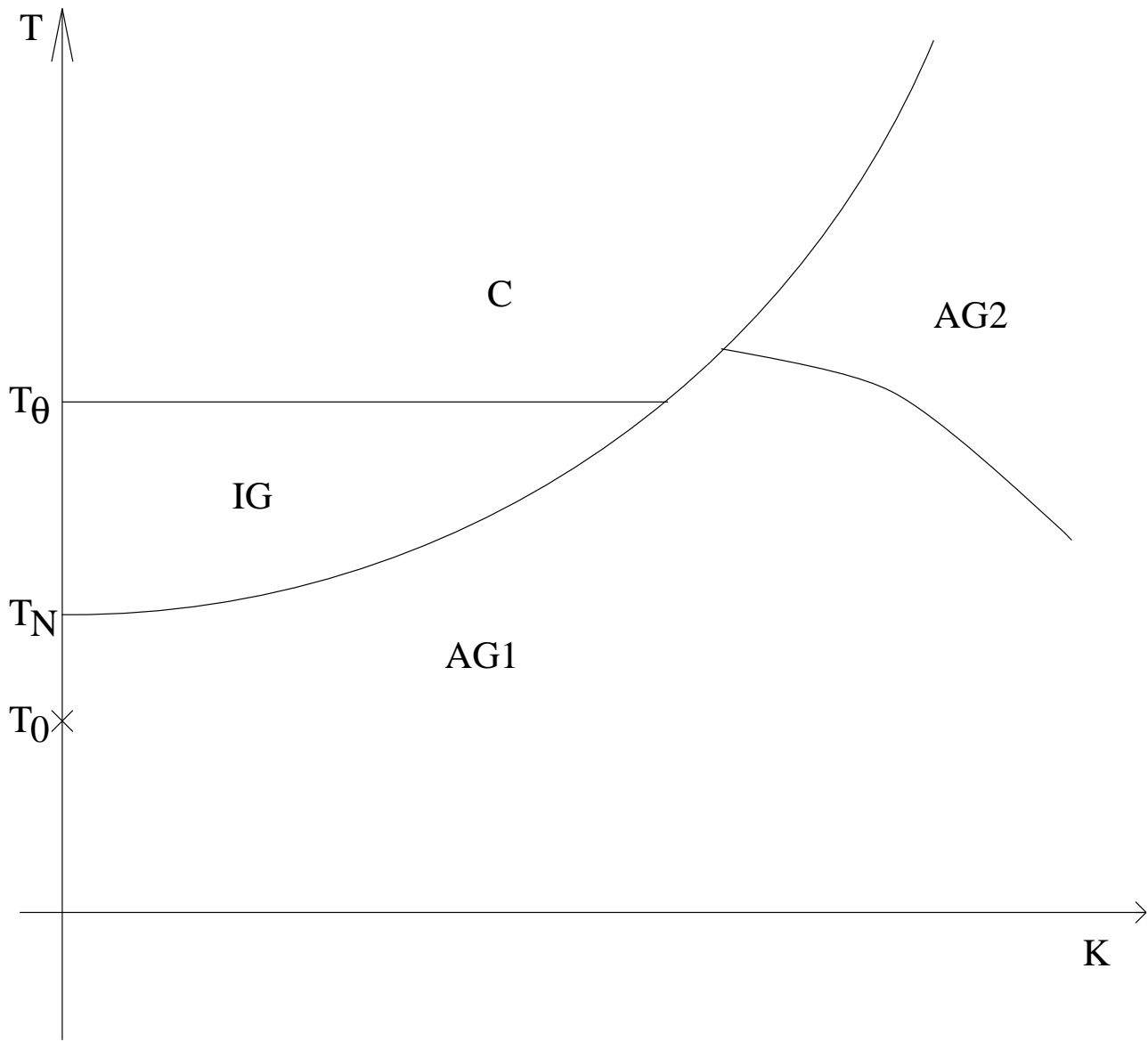
\end{document}